\documentclass[superscriptaddress,amsfonts,amssymb,amsmath,twocolumn,floatfix,prd]{revtex4}

\usepackage{soul}
\usepackage{ulem}
\usepackage{xcolor}
\usepackage{slashed}
\usepackage[compat=1.1.0]{tikz-feynman}
\usepackage{amsfonts} 
\usepackage[makeroom]{cancel}
\usepackage{txfonts}

\usepackage[USenglish]{babel}
\usepackage{epsfig}
\usepackage{graphics}
\usepackage{graphicx} 
\usepackage{amsmath}
\usepackage{MnSymbol}
\usepackage{color}
\usepackage{mathrsfs}
\usepackage{braket}
\usepackage{comment}
\usepackage{slashed}
\usepackage{float}
\usepackage{siunitx}
\usepackage{hyperref}

\hypersetup{
    colorlinks=true,
    linkcolor=blue,
    citecolor=blue,
    filecolor=magenta,
    urlcolor=cyan,
    pdfpagemode=FullScreen,
    }

\date{\today}

\begin{document}

\title{Emission of pairs of Minkowski photons through the lens of the Unruh effect}

\author{Juan V. O. Pêgas}
\email{jv.pegas@unesp.br}
\affiliation{Universidade Estadual Paulista, Instituto de Física Teórica, Rua Dr. Bento Teobaldo Ferraz 271, Bl. II, 01140-070, Sao Paulo, Sao Paulo, Brazil}%

\author{Robert Bingham}
\email{bob.bingham@stfc.ac.uk}
\affiliation{Rutherford Appleton Laboratory, Chilton, Didcot, Oxon OX11OQX, UK}
\affiliation{Department of Physics, University of Strathclyde, Glasgow G40NG, UK}
\author{Gianluca Gregori}
\email{gianluca.gregori@physics.ox.ac.uk}
\affiliation{Department of Physics, University of Oxford, Parks Road, Oxford OX13PU, UK}
\author{George E. A. Matsas}
\email{george.matsas@unesp.br}
\affiliation{Universidade Estadual Paulista, Instituto de Física Teórica, Rua Dr. Bento Teobaldo Ferraz 271, Bl. II, 01140-070, Sao Paulo, Sao Paulo, Brazil}%


\begin{abstract}
We discuss the emission of pairs of photons by charges with generic worldlines in the Minkowski vacuum from the viewpoint of inertial observers and interpret them from the perspective of Rindler observers.  We show that the emission of pairs of Minkowski photons—commonly referred to as Unruh radiation—corresponds, in general, to three distinct processes according to Rindler observers: scattering, and emission and absorption of pairs of Rindler photons. In the special case of uniformly accelerated charges, the radiation observed in the inertial frame can be fully described by the scattering channel in the Rindler frame. Therefore, the emission of pairs of Minkowski photons can be seen as further evidence supporting the Unruh effect. 
\end{abstract}

\maketitle

     \section{Introduction}
     \label{introduction}

One of the most paradigmatic effects of quantum field theory is the Unruh effect~\cite{U76}, which states that the usual inertial vacuum of Minkowski spacetime is perceived as a thermal bath of particles, with a temperature of
\begin{equation}
    T_U=\frac{\hbar a}{2\pi ck_B},
\end{equation}
by uniformly accelerated observers. Although the Unruh effect is necessary to maintain the consistency of quantum field theory in uniformly accelerated frames~\cite{UW84}---and, as such, requires no more experimental confirmation than standard free quantum field theory---its existence is often challenged (see, for instance, Refs.~\cite{Belinskii97,Narozhny01,Ford06,Gelfer15,Cruz16,OConnell20,Popruzhenko23}). As a consequence, much effort has been spent on proposals for experimentally observing the Unruh effect (see, \textit{e.g.}, Refs.~\cite{Bell83,Singleton07, VM01, Tajima99, V19, Lynch21,Leonhardt18,Matsas20,Bob08,Hessami25}). By observing the Unruh effect, \textit{we mean looking for signals in the laboratory frame that can be understood in the Rindler frame by taking into account the Unruh thermal bath}. Larmor radiation emitted by uniformly accelerated charges consists of a simple example of it~\cite{HMS92,HMS92b,MFL19,AMV25, P24,Vaccalis24}: in the inertial frame, each photon emitted by a uniformly accelerated charge corresponds, in the co-accelerated frame, to either the emission or absorption of a zero-energy Rindler photon from the thermal bath. These zero-energy Rindler photons, $\omega=0$, with non-zero transverse momenta, $k_\perp\neq 0$ (note that there is no dispersion relation connecting $\omega$ and $k_\perp$), are as well-defined as, say, Minkowski photons with $k_x=0$ and $k_\perp \neq 0$. However, the unfamiliar nature of these zero-energy modes has raised concerns about whether Larmor radiation can be interpreted as a signature of the Unruh effect~\cite{Narozhny01}. To circumvent this issue, Ref.~\cite{CLMV17} considered a non-uniformly accelerated charge to verify the existence of the Unruh thermal bath encoded in Larmor radiation, where, now, zero-energy Rindler photons no longer play a central role. Nevertheless, doubts and misconceptions regarding the interpretation of this proposal have persisted (see, for instance, Ref.~\cite{S17}).

In a distinct front, Refs.~\cite{S06,S09,S08, G24} have proposed using high-performance lasers to investigate higher-order effects in QED triggered by accelerated electrons. These effects could be interpreted in the framework of the Unruh effect, and their detection could therefore be seen as further evidence supporting the Unruh thermal bath. In particular, Refs.~\cite {S06,S09,S08} have analyzed the emission of pairs of Minkowski photons, which, according to the authors, would correspond, in the uniformly accelerated frame, to the scattering of Rindler photons by the electron. Moreover, this radiation would be distinguishable from Larmor radiation and could be directly detected in the laboratory. This perspective would gain relevance in light of the enormous progress in high-intensity laser technology, where lasers with intensities $\gtrsim 10^{19}~\text{W/cm}^2$ would accelerate electrons to an Unruh temperature $\gtrsim 1~\text{eV}$~\cite{S08}.

In this paper, we consider electric charges with classical worldlines emitting pairs of Minkowski photons as described by inertial observers, and discuss this process from the viewpoint of Rindler observers. Our results can be applied to physical situations where the radiation emission is dominated by soft photons, allowing us to disregard radiation reaction on the charge. (By {\it soft photons} we mean photons with energies much smaller than the electron mass.) Here, we explicitly show that the emission of pairs of soft photons in the inertial frame by a \textit{uniformly accelerated charge} corresponds in the Rindler frame to a Thomson scattering process of Rindler photons from the Unruh thermal bath, as conjectured in Refs.~\cite{S09,S08}. For charges undergoing \textit{non-uniform acceleration}, however, the corresponding description in the Rindler frame must be supplemented by two additional processes, namely the absorption and emission of pairs of Rindler photons. 

This paper is organized as follows. In Sec.~\ref{Scalar case} we begin reviewing the Rindler and Unruh modes for the scalar field. We then proceed analyzing the corresponding modes for the electromagnetic field. In Sec.~\ref{sec: scalar pairs case}, as a first step, we examine the emission of pairs of massless scalar particles in the Minkowski vacuum by an accelerated source in the inertial frame and its interpretation in the Rindler frame. Building on this analysis, in Sec~\ref{sec:EM case} we extend the previous discussion to the electromagnetic case. In Sec.~\ref{sec: Final} we summarize our conclusions. Hereafter, we assume metric signature $(+,-,-,-)$ and $k_B=c=\hbar=1$.

   \section{Rindler and Unruh modes }
   \label{Scalar case}
   
We begin with a brief review of the Rindler and Unruh modes for the scalar and electromagnetic fields. We address to Ref.~\cite{M08} for more details.

   \subsection{Scalar case}
   
   The free massless scalar field is described by the Lagrangian
   \begin{equation}
   \label{KG equation}
       \mathcal{L}=\frac{1}{2}\sqrt{-g}\nabla_\mu\phi\nabla^\mu\phi,
   \end{equation}
    where $g$ is the determinant of the metric. The corresponding field-operator solutions can be expanded in terms of plane waves as
   \begin{equation}
   \label{scalar inertial decomposition}
       \hat{\phi}=\int d^3k\left(\hat{a}^M_{\textbf{k}}f_{\textbf{k}}+\hat{a}^{M\dagger}_{\textbf{k}}\overline{f_{\textbf{k}}} \right),
   \end{equation}
where
    \begin{equation}
    f_\textbf{k} = \left[(2\pi)^32k\right]^{-1/2} e^{-ikt+ik_zz+i\textbf{k}_\perp\cdot\textbf{x}_\perp}.
    \end{equation}
Here, $(t,z,\textbf{x}_\perp)$ with $\textbf{x}_\perp \equiv(x,y)$ are usual inertial coordinates, $\textbf{k} \equiv (k_z,\textbf{k}_\perp)$, and $k\equiv||\textbf{k}||$.
The annihilation and creation operators $\hat{a}^M_{\textbf{k}}$ and $\hat{a}^{M\dagger}_{\textbf{k}}$ satisfy
   \begin{equation}
       \left[\hat{a}^M_{\textbf{k}},\hat{a}^{M\dagger}_{\textbf{k}'}\right]=\delta^{(3)}(\textbf{k}-\textbf{k}')
   \end{equation}
   with all other commutation relations vanishing. The Minkowski vacuum, $\ket{0_M}$, is defined by
   \begin{equation}
       \hat{a}_\textbf{k}^M\ket{0_M}=0,\quad \text{for all $\textbf{k}$}.
   \end{equation}

   In order to obtain the Rindler modes at the right Rindler wedge, $z>|t|$, it is convenient to use Rindler coordinates
   \begin{equation}
   \label{rindler coordinates}
       t=\frac{e^{a\xi}}{a}\sinh{(a\tau)},\quad
       z=\frac{e^{a\xi}}{a}\cosh{(a\tau)},
   \end{equation}
with which the line element takes the form
   
\begin{equation}
\label{line element}
    ds^2 = e^{2a\xi} \left(d\tau^2-d\xi^2\right)-dx^2-dy^2.
\end{equation}
Similarly, one can define Rindler coordinates $(\tilde{\tau},\tilde{\xi})$ covering the region $z<-|t|$, known as the left Rindler wedge, as
\begin{equation}
t=\frac{e^{a\Tilde{\xi}}}{a}\sinh{(a\Tilde{\tau})},\quad
       z=-\frac{e^{a\Tilde{\xi}}}{a}\cosh{(a\Tilde{\tau})},
\end{equation}
with which the line element takes the form of Eq.~\eqref{line element} with $\tau$ and $\xi$ replaced by $\Tilde{\tau}$ and $\Tilde{\xi}$.

Analogously to Eq.~\eqref{scalar inertial decomposition}, we can expand the scalar field in terms of left and right Rindler modes as 
\begin{equation}
    \hat{\phi} = \hat{\phi}_R+\hat{\phi}_L,
\end{equation}
where $\hat{\phi}_R$ is the scalar field restricted to the right Rindler wedge, given by
\begin{equation}
    \hat{\phi}_R=\int d^2\textbf{k}_\perp \int_0^\infty d\omega \left[\hat{a}^R_{\omega,\textbf{k}_\perp}v^R_{\omega,\textbf{k}_\perp}+\hat{a}^{R\dagger}_{\omega,\textbf{k}_\perp}\overline{v^R_{\omega,\textbf{k}_\perp}}\right].
\end{equation}
Here, the right Rindler modes are 
\begin{equation}
\label{Eq: scalar Rindler modes}    v^R_{\omega,\textbf{k}_\perp} = F_{\omega,\textbf{k}_\perp}(\xi,\textbf{x}_\perp)e^{-i\omega\tau}
\end{equation}
with
\begin{equation}
\label{function F}
    F_{\omega,\textbf{k}_\perp}(\xi,\textbf{x}_\perp)=\sqrt{\frac{\sinh{(\pi\omega/a)}}{4\pi^4a}}K_{i\omega/a}\left(\frac{k_\perp e^{a\xi}}{a}\right)e^{i\textbf{k}_\perp\cdot\textbf{x}_\perp},
\end{equation}
where $K_\nu(x)$ is the modified Bessel function of the second kind. The annihilation and creation operators $\hat{a}^R_{\omega,\textbf{k}_\perp}$ and $\hat{a}_{\omega,\textbf{k}_\perp}^{R\dagger}$ satisfy
\begin{equation}
    \left[\hat{a}^R_{\omega,\textbf{k}_\perp},\hat{a}^{R\dagger}_{\omega',\textbf{k}_\perp'}\right]=\delta(\omega-\omega')\delta^{(2)}(\textbf{k}_\perp-\textbf{k}'_\perp),
\end{equation}
with all other commutation relation vanishing. Similarly, the scalar field restricted to the left Rindler wedge, $\hat{\phi}_L$, is given by 
\begin{equation}
   \hat{\phi}_L=\int d^2\textbf{k}_\perp \int_0^\infty d\omega \left[\hat{a}^L_{\omega,\textbf{k}_\perp}v^L_{\omega,\textbf{k}_\perp}+\hat{a}^{L\dagger}_{\omega,\textbf{k}_\perp}\overline{v^L_{\omega,\textbf{k}_\perp}}\right],
\end{equation}
where the modes $v^L_{\omega, \textbf{k}_\perp}$ are obtained from $v_{\omega,\textbf{k}_\perp}^R$ by replacing $\tau$ and $\xi$ by $\Tilde{\tau}$ and $\Tilde{\xi}$, respectively. The annihilation and creation operators $\hat{a}^L_{\omega,\textbf{k}_\perp}$ and $\hat{a}_{\omega,\textbf{k}_\perp}^{L\dagger}$ satisfy
\begin{equation}
    \left[\hat{a}^L_{\omega,\textbf{k}_\perp},\hat{a}^{L\dagger}_{\omega',\textbf{k}_\perp'}\right]=\delta(\omega-\omega')\delta^{(2)}(\textbf{k}_\perp-\textbf{k}'_\perp),
\end{equation}
with all other commutation relations vanishing. The Fulling vacuum $\ket{0_F}$ is defined by requiring that $\hat{a}^R_{\omega,\textbf{k}_\perp}\ket{0_F}=\hat{a}^L_{\omega,\textbf{k}_\perp}\ket{0_F}=0$ for all $\omega$ and $\textbf{k}_\perp$.

Then, the ``\, $\mp$ \,'' Unruh modes are defined as
\begin{equation}
\label{Unruh mode - scalar}
    w_{(-,~\omega,\textbf{k}_\perp)}=\frac{v^R_{\omega,\textbf{k}_\perp}+e^{-\pi\omega/a}\overline{v^L_{\omega,-\textbf{k}_\perp}}}{\sqrt{1-e^{-2\pi\omega/a}}},
\end{equation}
and
\begin{equation}
\label{Unruh mode + scalar}
    w_{(+,~\omega,\textbf{k}_\perp)}=\frac{v^L_{\omega,\textbf{k}_\perp}+e^{-\pi\omega/a}\overline{v^R_{\omega,-\textbf{k}_\perp}}}{\sqrt{1-e^{-2\pi\omega/a}}}.
\end{equation}
The expansion of the scalar field in terms of the Unruh modes is given by
\begin{eqnarray}
\label{Expansion using Unruh modes for KG}
        \hat{\phi} &=& \int d^2\textbf{k}_\perp\int_0^{\infty}d\omega \left[\hat{a}_{(-,~\omega,\textbf{k}_\perp)}w_{(-,~\omega,\textbf{k}_\perp)} \right.
        \nonumber \\
        && \left.+\hat{a}_{(+,~\omega,\textbf{k}_\perp)}w_{(+,~\omega,\textbf{k}_\perp)}
        +\rm{H.c.}\right],
\end{eqnarray}
where $\rm{H.c}$ stands for Hermitian conjugate and 
\begin{equation}
    \left[\hat{a}_{(\pm,~\omega,\textbf{k}_\perp)},\hat{a}^\dagger_{(\pm,~\omega',\textbf{k}_\perp')}\right]=\delta(\omega-\omega')\delta^{(2)}(\textbf{k}_\perp-\textbf{k}'_\perp),
\end{equation}
with all other commutation relations vanishing. One can show that the Unruh annihilation operators $\hat{a}_{(\pm,~\omega,\textbf{k}_\perp)}$ are a combination of the Minkowski annihilation ones $\hat{a}^M_{\textbf{k}}$(see, \textit{e.g.},~\cite{AMV25}), and, thus, satisfy $\hat{a}_{(\pm,~\omega,\textbf{k}_\perp)}\ket{0_M}=0$, for all $\omega$ and $\textbf{k}_\perp$.

\subsection{Electromagnetic case}

Next, let us define the Unruh and Rindler modes for the electromagnetic field $\hat{A}_\mu$ described by the Lagrangian
\begin{equation}
\label{EM lagrangian}
    \mathcal{L} = -\frac{1}{4}\sqrt{-g}F_{\mu\nu}F^{\mu\nu}-\frac{1}{2}\sqrt{-g}\left(\nabla_\alpha A^\alpha\right)^2,
\end{equation}
where $F_{\mu\nu}=\nabla_\mu A_\nu-\nabla_\nu A_\mu$ and the last term is a gauge fixing term. The quantized electromagnetic field $\hat{A}_\mu$ can be expanded in terms of plane waves as
\begin{equation}
\label{Minkowski decomposition}
    \hat{A}_\mu = \int \frac{d^3k}{\sqrt{2(2\pi)^3k}}\sum_{\lambda=1}^2\left[\hat{a}^M_{(\lambda,\textbf{k})}\epsilon_\mu (\lambda,\textbf{k})e^{-ik_\mu x^\mu} + \rm{H.c.}   \right], 
\end{equation}
where $\lambda=1,2$ labels the physical linear polarizations and $\epsilon_\mu(\lambda,\textbf{k})$ are linear polarization vectors. The operators $\hat{a}^M_{(\lambda,\textbf{k})}$ and $\hat{a}_{(\lambda,\textbf{k})}^{M\dagger}$, for $\lambda=1,2$, satisfy
\begin{equation}
    \left[\hat{a}^M_{(\lambda,\textbf{k})},\hat{a}_{(\lambda',\textbf{k}')}^{M\dagger}\right]=\delta_{\lambda\lambda'}\delta^{(3)}(\textbf{k}-\textbf{k}'),
\end{equation}
with all other commutation relations vanishing. The Minkowski vacuum state for the electromagnetic field is defined by requiring $\hat{a}^M_{(\lambda, \textbf{k})}\ket{0_M}=0$ for all $\lambda$ and $\textbf{k}$.

In order to obtain the Rindler modes for the electromagnetic field it is convenient to use the Rindler coordinates~\eqref{rindler coordinates}. Thus, the electromagnetic field restricted to the right Rindler wedge, $\hat{A}^R_\mu$, can be decomposed as
\begin{equation}
    \hat{A}_\mu^R = \int d^2\textbf{k}_\perp\int_0^{\infty} d\omega \sum_{P=1}^{2}\left[\hat{a}^R_{(P,\omega,\textbf{k}_\perp)}A_\mu^{R(P,\omega,\textbf{k}_\perp)}+\rm{H.c.}\right],
\end{equation}
where $P$ labels the physical polarizations of the Rindler modes,   
\begin{equation}
\label{Rindler mode 1}
    A_\mu^{R(1,\omega,\textbf{k}_\perp)} = k_\perp^{-1}(0,0,k_yv^R_{\omega,\textbf{k}_\perp},-k_xv^R_{\omega,\textbf{k}_\perp}),
\end{equation}
\begin{equation}
\label{Rindler mode 2}
    A_\mu^{R(2,\omega,\textbf{k}_\perp)} = k_\perp^{-1}(\partial_\xi v^R_{\omega,\textbf{k}_\perp},\partial_\tau v^R_{\omega,\textbf{k}_\perp},0,0),
\end{equation}
and $v_{\omega,\textbf{k}_\perp}^R$ is given by Eq.~\eqref{Eq: scalar Rindler modes}. The annihilation and creation operators $\hat{a}^R_{(P,\omega,\textbf{k}_\perp)}$ and $\hat{a}^{R\dagger}_{(P,\omega,\textbf{k}_\perp)}$ satisfy
\begin{equation}
    \left[\hat{a}^R_{(P,\omega,\textbf{k}_\perp)},\hat{a}^{R^\dagger}_{(P',\omega',\textbf{k}'_\perp)}\right]=\delta_{PP'}\delta(\omega-\omega')\delta^{(2)}(\textbf{k}_\perp-\textbf{k}'_\perp),
\end{equation}
with all other commutation relations vanishing. The right Rindler vacuum for the electromagnetic field, $\ket{0_R}$, is defined by $\hat{a}^R_{(P,\omega,\textbf{k}_\perp)}\ket{0_R}=0$ for all $P,~\omega,$ and $\textbf{k}_\perp$. Similarly, the left Rindler modes, $A_\mu^{L(P,\omega,\textbf{k}_\perp)}$, are obtained from Eqs.~\eqref{Rindler mode 1} and \eqref{Rindler mode 2} by replacing $v_{\omega,\textbf{k}_\perp}^R$ by $v_{\omega,\textbf{k}_\perp}^L$, where the latter is obtained from the former by doing $(\tau,\xi)\to(\Tilde{\tau},\Tilde{\xi})$.

The ``$\mp$'' Unruh modes are then defined as~\cite{P24}
\begin{equation}
\label{Unruh modes}
    W_\mu^{(-,P,\omega,\textbf{k}_\perp)}=\frac{A_\mu^{R(P,\omega,\textbf{k}_\perp)}+e^{-\pi\omega/a}\overline{A^{L(P,\omega,-\textbf{k}_\perp)}_\mu}}{\sqrt{1-e^{-2\pi\omega/a}}}
\end{equation}
and
\begin{equation}
\label{Unruh modes 2}    W_\mu^{(+,P,\omega,\textbf{k}_\perp)}=\frac{A_\mu^{L(P,\omega,\textbf{k}_\perp)}+e^{-\pi\omega/a}\overline{A^{R(P,\omega,-\textbf{k}_\perp)}_\mu}}{\sqrt{1-e^{-2\pi\omega/a}}}.
\end{equation}
The expansion of the electromagnetic field in terms of the Unruh modes above is given by
\begin{eqnarray}
\label{Expansion using Unruh modes}
        \hat{A}_\mu &=& \int d^2\textbf{k}_\perp\int_0^{\infty}d\omega \sum_{P=1}^{2}\left[\hat{a}_{(-,P,\omega,\textbf{k}_\perp)}W_\mu^{(-,P,\omega,\textbf{k}_\perp)} \right.
        \nonumber \\
        &&\left.+\hat{a}_{(+,P,\omega,\textbf{k}_\perp)}W_\mu^{(+,P,\omega,\textbf{k}_\perp)}
        +\rm{H.c.}\right]
\end{eqnarray}
with
\begin{equation}
    \left[\hat{a}_{(\pm,P,\omega,\textbf{k}_\perp)},\hat{a}^\dagger_{(\pm,P',\omega',\textbf{k}_\perp')}\right]=\delta_{PP'}\delta(\omega-\omega')\delta^{(2)}(\textbf{k}_\perp-\textbf{k}'_\perp),
\end{equation}
and all other commutation relations vanish. As in the scalar case, the Unruh annihilation operators $\hat{a}_{(\pm,P,\omega,\textbf{k}_\perp)}$ are a combination of the Minkowski ones $\hat{a}^M_{(\lambda,\textbf{k})}$~\cite{Vaccalis24}, and, thus, $\hat{a}_{(\pm,P,\omega,\textbf{k}_\perp)}\ket{0_M}=0$, for all $P,\omega,\textbf{k}_\perp$.

\section{Pair production of massless scalar particles and the Unruh thermal bath}
\label{sec: scalar pairs case}

Our ultimate goal is to interpret the emission of pairs of Minkowski photons by accelerated charges in the Rindler frame. For this purpose, we shall use the effective interaction action 
$$
\hat{S}_{I}=-\int d^4x\sqrt{-g}~j(x):\hat{A}_\mu(x)\hat{A}^\mu(x):.
$$ 
Nevertheless, as a first step, let us consider the analogous problem of an accelerated scalar source $j(x)$ emitting pairs of massless scalar particles as given by  
\begin{equation}
\label{scalar action}
    \hat{S}_I = -\int d^4x \sqrt{-g} \; j(x):\hat{\phi}(x)^2:, 
\end{equation}
where $``:\quad:"$ indicates normal ordering. 

\subsection{General sources}

At first order in perturbation theory, the probability of emission of a pair of scalar Minkowski photons $\ket{\textbf{k};\textbf{k}'}$ with three-momenta $\textbf{k}$ and $\textbf{k}'$  is
\begin{equation}
\label{Pm scalar}
        P_M^S = \int d^3k \int d^3k'\left|\bra{\textbf{k};\textbf{k}'} \hat{S}_I\ket{0_M}\right|^2,
\end{equation}
where the $S$ label stands for ``scalar''. This can be recast as
\begin{equation}
\label{PM scalar rewritten}
    P_M^S=\bra{f} \left(\int d^3k \int d^3k'\ket{\textbf{k};\textbf{k}'}\bra{\textbf{k};\textbf{k}'}\right) \ket{f},
\end{equation}
where 
\begin{equation}
\label{scalar two state initial}
    \ket{f}\equiv -i\int d^4x\sqrt{-g}~j(x):\hat{\phi}(x)^2:\ket{0_M}.
\end{equation}
Using in Eq.~\eqref{PM scalar rewritten} the decomposition of the identity operator
\vskip 0.7 cm
\begin{equation}
\label{identity operator scalar}
    I=\int d^3k\ket{\textbf{k}}\bra{\textbf{k}}+\frac{1}{2!}\int d^3k \int d^3k'\ket{\textbf{k};\textbf{k}'}\bra{\textbf{k};\textbf{k}'}+\ldots,
\end{equation}
we get
\begin{equation}
\label{PM and inner product scalar}
        P_M^S=2\braket{f|f}.
\end{equation}
Now, to evaluate Eq.~\eqref{PM and inner product scalar}, we use the decomposition of $\hat{\phi}$ in terms of the Unruh modes~\eqref{Expansion using Unruh modes for KG} to calculate Eq.~\eqref{scalar two state initial}, obtaining
\begin{widetext}
\begin{eqnarray}
    \label{inner product state 2 scalar}
    \braket{f|f} 
    &=&   2 \int_{_\mathcal{I}} d\mu  \int_{_{\mathcal{I}'}} d\mu'
    \int d^4x\sqrt{-g}~j(x)\int d^4x'\sqrt{-g'}~j(x')
    \left[ w_{(-,~\omega,k_\perp)}(x)w_{(-,~\omega',k'_\perp)}(x)\overline{w_{(-,~\omega, k_\perp)}(x')}~\overline{w_{(-,~\omega',k'_\perp)}(x')}\right. \nonumber \\
    &+& w_{(-,~\omega, k_\perp)}(x)w_{(+,~\omega',k'_\perp)}(x)\overline{w_{(-,~\omega, k_\perp)}(x')}~\overline{w_{(+,~\omega',k'_\perp)}(x')} 
    + w_{(+,~\omega, k_\perp)}(x)w_{(-,~\omega', k'_\perp)}(x)\overline{w_{(+,~\omega, k_\perp)}(x')}~\overline{w_{(-,~\omega',k'_\perp)}(x')} \nonumber \\
     &+&  \left. w_{(+,~\omega, k_\perp)}(x)w_{(+,~\omega',k'_\perp)}(x)\overline{w_{(+,~\omega, k_\perp)}(x')}~\overline{w_{(+,~\omega',k'_\perp)}(x')}
        \right] 
\end{eqnarray}
where $\mathcal{I}\equiv\{\omega,\textbf{k}_\perp\}$ and $d\mu\equiv d\omega d^2\textbf{k}_\perp$ (and similarly for $\mathcal{I}'$ and $d\mu'$). 
Next, let us assume that the current $j(x)$ has support in the right Rindler wedge (where we recall that $v^L_{\omega,\textbf{k}_\perp}=0$). In this case, using the Unruh modes~\eqref{Unruh mode - scalar} and \eqref{Unruh mode + scalar}, the emission probability~\eqref{PM and inner product scalar} reads
\begin{eqnarray}
    \label{two-photon probability scalar case}
    \!\!\!\!\!\!\!\!\!\! && P_M^S = \! \int_{_\mathcal{I}} \! d\mu \! \int_{_{\mathcal{I}'}} \! d\mu'~n(\omega)n(\omega')\left|2 \int  d^4x\sqrt{-g}~j~ v^R_{\omega',k'_\perp} v^R_{\omega,k_\perp}\right|^2 \! + \! \int_{_\mathcal{I}}  d\mu\int_{_{\mathcal{I}'}} d\mu'~n(\omega)\left[1+n(\omega')\right]\left|2 \int  d^4x\sqrt{-g}~j~v^R_{\omega, k_\perp} ~\overline{v^R_{\omega',k'_\perp}}\right|^2 \nonumber \\
    \!\!\!\!\!\!\!\!\!\! && + \!\!\int_{_\mathcal{I}}\!\! d\mu \!\!\int_{_{\mathcal{I}'}} \!\! d\mu'~n(\omega')\left[1+n(\omega)\right]\left|2 \!\! \int \!\!d^4x\sqrt{-g}~j~v^R_{\omega',k'_\perp} ~\overline{v^R_{\omega, k_\perp}}\right|^2 \!\! + \!\! \int_{_\mathcal{I}} \!\! d\mu \!\! \int_{_{\mathcal{I}'}} \!\! d\mu' \left[1+n(\omega)\right] \left[1+n(\omega')\right] \left|2 \!\! \int \!\! d^4x\sqrt{-g}~j~\overline{v^R_{\omega',k'_\perp}}~\overline{v^R_{\omega, k_\perp}}\right|^2
\end{eqnarray}
where $n(\omega)\equiv 1/(e^{2\pi\omega/a}-1)$.
\end{widetext}
Now, to interpret the emission of the pair of Minkowski scalar photons obtained above from the perspective of Rindler observers, we define the amplitudes corresponding to the absorption and emission of two Rindler scalar particles, as well as to the scattering of a Rindler scalar particle by the source:
\vspace{-0.23em}
\begin{eqnarray}
        \label{scalar absorption amplitude}{}^R\mathscr{A}^{S,\text{abs}}_{\omega, k_\perp;\omega',k'_\perp}&\equiv& i\bra{0_R}\hat{S}_I\ket{\omega, k_\perp;\omega',k'_\perp}\nonumber \\
        &=&-2i\int d^4x\sqrt{-g}~j~v^R_{\omega',k'_\perp} v^R_{\omega,k_\perp},
    \end{eqnarray}
\begin{eqnarray}
\label{scalar emission amplitude}
        {}^R\mathscr{A}^{S,\text{em}}_{\omega, k_\perp;\omega',k'_\perp}&\equiv& i\bra{\omega, k_\perp;\omega',k'_\perp}\hat{S}_I\ket{0_R}\nonumber \\
        &=&-2i\int d^4x\sqrt{-g}~j~\overline{v^R_{\omega',k'_\perp}}~\overline{v^R_{\omega,k_\perp}},
\end{eqnarray}
\begin{eqnarray}
\label{scalar scattering amplitude}
        {}^R\mathscr{A}^{S,\text{scatt}}_{\omega, k_\perp;\omega',k'_\perp}&\equiv& i\bra{\omega, k_\perp}\hat{S}_I\ket{\omega',k'_\perp}\nonumber \\
        &=&-2i\int d^4x\sqrt{-g}~j~v^R_{\omega',k'_\perp}~\overline{v^R_{\omega,k_\perp}}.
\end{eqnarray}
Using Eqs.~\eqref{scalar absorption amplitude}, \eqref{scalar emission amplitude}, and \eqref{scalar scattering amplitude} in Eq.~\eqref{two-photon probability scalar case}, one has
\begin{eqnarray}
\label{scalar inertial probability}
        P_M^S &=& \int_{_\mathcal{I}} d\mu\int_{_{\mathcal{I}'}} d\mu'
        \left[1+n(\omega)\right]\left[1+n(\omega')\right]\left|{}^R\mathscr{A}^{S,\text{em}}_{\omega, k_\perp;\omega',k'_\perp}\right|^2
        \nonumber \\
        &+&\int_{_\mathcal{I}} d\mu\int_{_{\mathcal{I}'}} d\mu'\; n(\omega) \; n(\omega') \; \left|{}^R\mathscr{A}^{S,\text{abs}}_{\omega, k_\perp;\omega',k'_\perp}\right|^2
        \nonumber \\
        &+&\int_{_\mathcal{I}} d\mu\int_{_{\mathcal{I}'}} d\mu'~n(\omega)\left[1+n(\omega')\right]\left|{}^R\mathscr{A}^{S,\text{scatt}}_{\omega',k'_\perp;\omega,k_\perp}\right|^2
        \nonumber \\
        &+&\int_{_\mathcal{I}} d\mu\int_{_{\mathcal{I}'}} d\mu'~n(\omega')\left[1+n(\omega)\right]\left|{}^R\mathscr{A}^{S,\text{scatt}}_{\omega, k_\perp;\omega',k'_\perp}\right|^2.
\end{eqnarray}

Thus, the emission of a pair of scalar particles with transverse momenta $k_\perp$ and  $k'_\perp$ in the usual vacuum of inertial observers [left-hand side of  Eq.~\eqref{scalar inertial probability}] should be associated in general to three processes in the Rindler frame: absorption (emission) of two Rindler scalar particles from (to) the Unruh thermal bath with the same transverse momenta $k_\perp$ and  $k'_\perp$, and scattering of a Rindler scalar particle from the Unruh thermal bath with $k_\perp$ to $k'_\perp$ (or $k'_\perp$ to $k_\perp$).

\subsection{Uniformly accelerated sources}

In the general case considered above, we have seen that the emission of pairs of Minkowski particles is perceived by uniformly accelerated observers as a combination of three distinct processes. We now turn our attention to the specific situation of a uniformly accelerated source. In this case, we shall see that only the scattering contribution survives out of the three. For this purpose, we will carry on independent calculations in the inertial and Rindler frames.

\subsubsection{Inertial frame calculation}

Let us consider that the source is accelerated along the z-direction with $\xi,x,y=0$. In this case, $j(x)=\mathfrak{g}\delta(\xi)\delta^{(2)}(\textbf{x}_\perp)$, where $\mathfrak{g}$ is a coupling constant. By using the plane-wave decomposition~\eqref{scalar inertial decomposition} to evaluate the Minkowski pair-emission amplitude at first order
\begin{eqnarray}
\label{Minkowski scalar emission amplitude}
        {}^M\mathscr{A}_{\textbf{k}\textbf{k}'}&\equiv& i\bra{\textbf{k};\textbf{k}'}\hat{S}_I\ket{0_M},
\end{eqnarray}
we get
\begin{equation}
    {}^M\mathscr{A}_{\textbf{k}\textbf{k}'}=-\frac{i\mathfrak{g}}{(2\pi)^3}\int d\tau \frac{e^{i(k+k')t(\tau)-i(k_z+k_z')z(\tau)}}{\sqrt{kk'}},
\end{equation}
where $t(\tau)$ and $z(\tau)$ are given by Eq.~\eqref{rindler coordinates} with $\xi=0$. Using this amplitude, the probability of emission
\begin{equation}
    P_M^S = \int d^3\textbf{k}\int d^3\textbf{k}' \left|{}^M\mathscr{A}_{\textbf{k}\textbf{k}'}\right|^2
\end{equation}
reads
\begin{eqnarray}
    P^S_M &=& \frac{\mathfrak{g}^2}{64\pi^6}\int_{-\infty}^{\infty} dT\int_{-\infty}^\infty d\sigma\int d^3\textbf{k}\int d^3\textbf{k}'\frac{1}{kk'} 
    \nonumber \\
    &\times& \exp{\left\{\frac{2i}{a}\sinh{(a\sigma/2)}\left[(k+k')\cosh{(aT)}\right]\right\}} 
    \nonumber \\
    &\times& \exp{\left\{-\frac{2i}{a}\sinh{(a\sigma/2)}\left[(k_z+k_z')\sinh{(aT)}\right]\right\}},
\end{eqnarray}
where $T\equiv (\tau+\tau')/2$ and $\sigma\equiv\tau-\tau'$. Applying the change of  coordinates~\cite{HMS92}
\begin{equation*}
        k_z=\Tilde{k}_z\cosh{(aT)}+\Tilde{k}\sinh{(aT)},
\end{equation*}
\begin{equation*}
        k'_z=\Tilde{k}'_z\cosh{(aT)}+\Tilde{k}'\sinh{(aT)},
\end{equation*}
where $\Tilde{k} \equiv \sqrt{\Tilde{k}_z+k_\perp^2}$ and similarly for $\Tilde{k}'$, the integrand becomes independent of $T$. Using that $d^3\Tilde{\textbf{k}}=\Tilde{k}^2\sin\theta\,d\Tilde{k}\,d\phi\, d\theta$ and integrating over the angles, we arrive at the total emission rate
\begin{equation}
\label{Eq: 49}
    \frac{P^S_M}{T_{\rm tot}} = \frac{\mathfrak{g}^2}{4\pi^4}\int_{-\infty}^{\infty} \!\!\!\! d\sigma \int_0^{\infty}\!\!\!\!  d\Tilde{k}\,\Tilde{k}\int_{0}^{\infty} \!\!\!\! d\Tilde{k}'\,\Tilde{k}'e^{(2i/a)(\Tilde{k}+\Tilde{k}')\sinh{(a\sigma/2)}},
\end{equation}
where $\int_{-\infty}^{\infty} dT\to T_{\rm tot}$ is the total proper time, and we recall that $T \equiv (\tau+\tau')/2$. We shall note that the total emission probability $P_M^S$ diverges since the source is accelerated from the past to the future infinity. Yet, the total emission rate ${P^S_M}/{T_{\rm tot}}$ has a well-defined finite value, as we will explicitly see ahead.
To proceed, let us define
$\lambda\equiv e^{a\sigma/2}$ to cast Eq.~\eqref{Eq: 49} as 
\begin{equation}
    \frac{P^S_M}{T_{\rm tot}} = \frac{\mathfrak{g}^2}{2\pi^4 a}\int_0^{\infty} \!\!\!\! d\Tilde{k}~\Tilde{k}\int_0^\infty \!\!\!\! d\Tilde{k}'~\Tilde{k}' \int_0^\infty \!\!\!\! d\lambda ~\lambda^{-1}e^{(i/a)(\Tilde{k}+\Tilde{k}')(\lambda-\lambda^{-1})}.
\end{equation}
Using~\cite{Gradz}
\begin{equation}
    \int_0^\infty dx x^{\nu-1}e^{-\frac{\beta}{x}-\gamma x}=2\left(\frac{\beta}{\gamma}\right)^{\nu/2}K_{\nu}\left(2\sqrt{\beta\gamma}\right),
\end{equation}
valid for $\rm{Re}(\beta)>0, \rm{Re}(\gamma)>0$, we can perform the $\lambda$ integral, obtaining
\begin{equation}
    \frac{P^S_M}{T_{\rm tot}} = \frac{\mathfrak{g}^2}{\pi^4a}\int_0^{\infty}d\Tilde{k}\,\Tilde{k}\int_0^\infty d\Tilde{k}' \,\Tilde{k}' K_0\left[\frac{2}{a}(\Tilde{k}+\Tilde{k}')\right].
\end{equation}
To perform the remaining integrals, let us define 
\begin{equation}
        u  \equiv (\Tilde{k}+\Tilde{k}')/2, \quad
        v  \equiv \Tilde{k}-\Tilde{k}',
\end{equation}
yielding
\begin{eqnarray}
\label{inertial density energy}
        \frac{P^S_M}{T_{\rm tot}}&=&\frac{\mathfrak{g}^2}{\pi^4a}\int_0^\infty du\,K_0\left(\frac{4u}{a}\right) \int_{-2u}^{2u}dv\left(u^2-\frac{v^2}{4}\right)
        \nonumber \\
        &=&\frac{8\mathfrak{g}^2}{3\pi^4a}\int_0^\infty u^3K_0\left(\frac{4u}{a}\right)du.
\end{eqnarray}
The last integral can be performed by noticing~\cite{Gradz}
\begin{equation}
    \int_0^\infty dx x^\mu K_\nu (ax)=2^{\mu-1}a^{-\mu-1}\Gamma\left(\frac{1+\mu+\nu}{2}\right)\Gamma\left(\frac{1+\mu-\nu}{2}\right)
\end{equation}
for $\mathrm{Re}(\mu+1\pm\nu)>0$ and $\mathrm{Re}(a)>0$. Therefore, the emission rate of pairs of massless scalar Minkowski particles is
\begin{equation}
\label{Eq: scalar inertial rate}
    \frac{P_M^S}{T_{\rm tot}}=\frac{\mathfrak{g}^2a^3}{24\pi^4}.
\end{equation}
 \subsubsection{Rindler frame calculation}
 Now, we shall evaluate the probability rate associated with the same uniformly accelerated current $j(x)$ according to Rindler observers by considering the Unruh thermal bath. The response would consist, in principle, of scattering, emission, and absorption of Rindler particles to and from the Unruh thermal bath. The corresponding amplitudes~\eqref{scalar absorption amplitude}, \eqref{scalar emission amplitude}, and \eqref{scalar scattering amplitude} are in this case
\begin{eqnarray}
    {}^R\mathscr{A}^{S,\text{em}}_{\omega, k_\perp;\omega',k'_\perp}
   \!\! &=& \!\! {}^R\mathscr{A}^{S,\text{abs}}_{\omega, k_\perp;\omega',k'_\perp}
    \nonumber \\
   \!\! &=& \!\! -4\pi i\mathfrak{g}F_{\omega, \textbf{k}_\perp}(0,\textbf{0})F_{\omega',\textbf{k}'_\perp}(0,\textbf{0})\delta(\omega+\omega')
\end{eqnarray}
and
\begin{equation}
\label{scattering amplitude uniformly accelerated case}
   {}^R\mathscr{A}^{S,\text{scatt}}_{\omega, k_\perp;\omega',k'_\perp} = -4\pi i\mathfrak{g}F_{\omega,\textbf{k}_\perp}(0,\textbf{0})F_{\omega,\textbf{k}'_\perp}(0,\textbf{0})\delta(\omega-\omega'),
\end{equation}
where we recall that $F_{\omega,\textbf{k}_\perp}(\xi,\textbf{x}_\perp)$ was defined in Eq.~\eqref{function F}.

The corresponding probabilities are given by integrating the square of the absolute value of the amplitudes with the corresponding thermal factors as in Eq.~\eqref{scalar inertial probability}. For the absorption process, we have
\begin{equation}
    \label{Eq: abs UA}P^{S,\text{abs}}_R=16\pi^2\mathfrak{g}^2\int_{_\mathcal{I}} d\mu\int_{_{\mathcal{I}'}} d\mu' f(\omega,\omega',\textbf{k}_\perp,\textbf{k}'_\perp)\delta^2(\omega+\omega'),
\end{equation}
where 
\begin{equation}
        f(\omega,\omega',\textbf{k}_\perp,\textbf{k}'_\perp) \equiv F_{\omega,\textbf{k}_\perp}(0,\textbf{0})^2F_{\omega',\textbf{k}'_\perp}(0,\textbf{0})^2n(\omega)n(\omega').
\end{equation}
Since both $\omega$ and $\omega'$ are positive, the delta function $\delta(\omega + \omega')$ vanishes Eq.~\eqref{Eq: abs UA}. The same reasoning applies to the absorption probability. In this way, for the uniformly accelerated particular case
\begin{equation}
    P_R^{S,\text{abs}}=P_R^{S,\text{em}}=0,
\end{equation}
and only the scattering process will contribute in the Rindler frame. Note that the above argument does not hold for non-uniformly accelerated sources, since no delta function $\delta(\omega+\omega')$ is present there; in that case, Rindler observers credit the external agent for providing the necessary work to account for the emission and absorption processes to occur.

Now, using Eq.~\eqref{scattering amplitude uniformly accelerated case}, the scattering probability reads
\begin{eqnarray}
    P_R^{S,\text{scatt}}
    &=&
    4\pi^2\mathfrak{g}^2 \int_{_\mathcal{I}} d\mu\int_{_{\mathcal{I}'}} d\mu' \frac{F_{\omega, \textbf{k}_\perp}(0,\textbf{0})^2F_{\omega,\textbf{k}'_\perp}(0,\textbf{0})^2}{\sinh^2{(\pi\omega/a)}}
    \nonumber \\
    &\times &
    \delta^2(\omega-\omega'),
\end{eqnarray}
where we have used 
$$
n(\omega)\left[1+n(\omega)\right]=(4\sinh^2{(\pi\omega/a)})^{-1}.
$$ 
Recalling that $d\mu'=d\omega'd^2\textbf{k}'_\perp$, we integrate over $\omega'$, getting
\begin{equation}
    \label{Equação 49}
    \frac{P_R^{S,\text{scatt}}}{T_{\rm tot}}=\frac{\mathfrak{g}^2}{8\pi^7a^2}\int_{_\mathcal{I}} d\mu\int d^2\textbf{k}'_\perp K_{i\omega/a}\left(\frac{k_\perp}{a}\right)^2K_{i\omega/a}\left(\frac{k'_\perp}{a}\right)^2,
\end{equation}
where we have used~(see, {\it e.g.,} Ref.~\cite{M08}) 
\begin{eqnarray}
T_{\rm tot} 
&=& 
\int_{-\infty}^{\infty} d\tau 
\nonumber \\
&=& 
2\pi \lim_{\omega\to 0} \frac{1}{2\pi}\int_{-\infty}^{\infty} d\tau  e^{i\omega \tau} 
\nonumber \\
&=& 2\pi\delta(0)
\end{eqnarray}
is the total proper time. Next, by using $d\mu=d\omega\, d^2\textbf{k}_\perp$ and $d^2\textbf{k}_\perp=k_\perp dk_\perp d\phi$, and integrating over the angles, we arrive at
\begin{equation}
\frac{P_R^{S,\text{scatt}}}{T_{\rm tot}}=\frac{\mathfrak{g}^2}{2\pi^5a^2}\int_0^\infty d\omega \left|\int_0^\infty dk_\perp k_\perp K_{i\omega/a}\left(\frac{k_\perp}{a}\right)^2\right|^2.
\end{equation}
The $k_\perp$ integral can be easily performed using~\cite{Gradz}
\begin{equation}
    \int_0^\infty x K_\nu(ax)K_\nu(bx) dx=\frac{\pi(ab)^{-\nu}\left(a^{2\nu}-b^{2\nu}\right)}{2\sin{(\nu\pi)}(a^2-b^2)},
\end{equation}
valid for $|\mathrm{Re}(\nu)|<1,~\mathrm{Re}(a+b)>0$, giving
\begin{eqnarray}      
\frac{P_R^{S,\text{scatt}}}{T_{\rm tot}}&=&\frac{\mathfrak{g}^2}{8\pi^3}\int_0^\infty d\omega \frac{\omega^2}{\sinh^2{(\pi\omega/a)}}\label{energy distribution Rindler}
\nonumber \\
&=&\frac{\mathfrak{g}^2a^3}{48\pi^4}
\label{scattered 1}.
\end{eqnarray}
By comparing Eqs.~\eqref{scattered 1} and~\eqref{Eq: scalar inertial rate}, we have 
\begin{equation}
\label{Eq: scatt=inertial}
P^{S}_M=2P^{S,\text{scatt}}_R,
\end{equation}
which is in agreement with
Eq.~\eqref{scalar inertial probability}. Thus, we have established by explicit calculation that the emission rate of pairs of Minkowski scalar particles from a uniformly accelerated source corresponds, in the co-accelerated frame, to the scattering of Rindler particles of the Unruh thermal bath.
\subsubsection{Energy spectrum in the inertial and Rindler frames}
\begin{figure}[h!]
    \centering
    \includegraphics[width=1\linewidth]{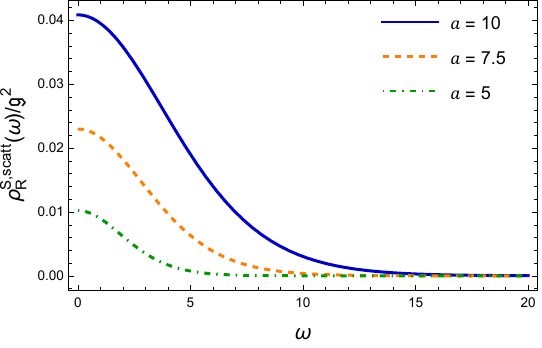}
    \caption{Energy distribution for scattered Rindler particles of the Unruh thermal bath for different values of the source's proper acceleration. }
    \label{fig:Rindler plot}
\end{figure}

\begin{figure}[h!]
    \centering
    \includegraphics[width=1\linewidth]{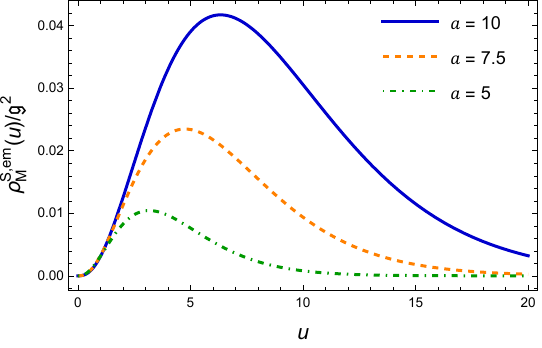}
    \caption{Energy distribution of the mean energy emitted in the inertial frame for different values of the source's proper acceleration. }
    \label{fig:inertial plot}
\end{figure}

As we have just established, the total response as calculated in the inertial and Rindler frames equal to each other (as it should be since this is a physical observable), although the corresponding interpretation is distinct in each frame. 
This is reinforced by studying the energy distribution of emission and scattering of Minkowski and Rindler particles, respectively. From Eq.~\eqref{energy distribution Rindler}, the energy distribution for scattered Rindler particles as a function of the particle's energy $\omega$ is (see Fig.~\ref{fig:Rindler plot})
\begin{equation}
\label{rho_omega}
    \rho^{S,\text{scatt}}_R(\omega)\equiv\frac{\mathfrak{g}^2}{8\pi^3}\frac{\omega^2}{\sinh^2{(\pi\omega/a)}}.
\end{equation}
Note that the scattering of soft Rindler particles is favored over high-frequency ones.
This is in contrast with what is calculated for the emission of pairs of Minkowski particles. Figure~\ref{fig:inertial plot} depicts the energy distribution 
\begin{equation}
    \label{rho_k}
    \rho^{S,\text{em}}_M(u)\equiv\frac{8\mathfrak{g}^2}{3\pi^4a}u^3K_0\left(\frac{4u}{a}\right)
\end{equation}
for the emitted pair of Minkowski particles as a function of the mean energy $u\equiv(k+k')/2$. The graph reveals a peak whose position depends on the proper acceleration of the source, with negligible contribution from soft and high-energy modes.
\section{Emission of pairs of Minkowski photons and the Unruh thermal bath}
\label{sec:EM case}
We now turn to the investigation of the emission of pairs of Minkowski photons and its interpretation in the Rindler frame. For this purpose, let us consider the \textit{effective} interaction action
\begin{equation}
\label{interaction action}
    \hat{S}_{I}=-\int d^4x\sqrt{-g}~j(x):\hat{A}_\mu(x)\hat{A}^\mu(x):
\end{equation}
in the radiation gauge, i.e., $A^0=0$ and $\mathbf{\nabla}\cdot\mathbf{A}=0$. We emphasize that we are interested here in the case where the energies of the emitted photons are much smaller than the electron’s rest mass $m$, allowing us to neglect back-reaction effects. This places us in a semiclassical regime, where the electron follows a well-defined trajectory, described by a classical ``scalar current'' $j(x)$, while the electromagnetic field, $\hat{A}_\mu$, is quantized. Note that $j(x) \propto e^2/m$, reflecting the fact that each  $e^- \, \gamma \, e^-$ vertex contributes to the amplitude with an elementary charge factor $e$, while the propagator scales with $1/m$ at low energies (see Appendix~\ref{Appendix} for more details). 

Specifically, we aim to evaluate the probability of emitting pairs of photons in the inertial frame and examine how this process is perceived in the Rindler frame. The probability of emission of a pair of Minkowski photons $\ket{\lambda,\textbf{k};\lambda',\textbf{k}'}$ with three-momenta $\textbf{k}$ and $\textbf{k}'$, and physical polarizations $\lambda$ and $\lambda'$, respectively, at first order in perturbation theory, is 
\begin{equation}
\label{Pm}
        P_M^{E} = \int d^3k \int d^3k'\sum_{\lambda}\sum_{\lambda'}\left|\bra{\lambda,\textbf{k};\lambda',\textbf{k}'} \hat{S}_I\ket{0_M}\right|^2,\\
\end{equation}
where the $E$ label stands for ``electromagnetic''. This  can be recast as (see Eq.~\eqref{scalar two state initial})
\begin{equation}
\label{PM rewritten}
    P_M^{E}=\bra{f} \left(\int d^3k \int d^3k'\sum_{\lambda}\sum_{\lambda'}\ket{\lambda,\textbf{k};\lambda',\textbf{k}'}\bra{\lambda,\textbf{k};\lambda',\textbf{k}'}\right) \ket{f},
\end{equation}
where
\begin{equation}
\label{two state initial}
    \ket{f}\equiv -i\int d^4x\sqrt{-g}~j(x):\hat{A}_\mu(x)\hat{A}^\mu(x):\ket{0_M}.
\end{equation}
Using in Eq.~\eqref{PM rewritten} the decomposition of the identity operator
\begin{eqnarray*}
\label{identity operator}
    I &=& \int d^3k\sum_{\lambda}\ket{\lambda,\textbf{k}}\bra{\lambda,\textbf{k}}
    \\
    &+&\frac{1}{2!}\int d^3k \int d^3k'\sum_{\lambda}\sum_{\lambda'}\ket{\lambda,\textbf{k};\lambda',\textbf{k}'} \bra{\lambda,\textbf{k};\lambda',\textbf{k}'}+\ldots,
\end{eqnarray*}
we get
\begin{equation}
\label{PM and inner product}
P_M^{E}=2\braket{f|f}.
\end{equation}
\vspace{0.5cm}
Now, to evaluate Eq.~\eqref{PM and inner product}, we use the decomposition of $\hat{A}_\mu$ in terms of the Unruh modes~\eqref{Expansion using Unruh modes} to calculate Eq.~\eqref{two state initial}, obtaining  
\begin{eqnarray}
\label{inner product state 2}
    &&\!\!\!\!\!\!\!\! \braket{f|f}
    =2\sumint_{_\mathcal{I}} \! d\mu\sumint_{_{\mathcal{I}'}} \! d\mu'
        \!\! \int \!\! d^4x\sqrt{-g(x)}~j(x)\!\! \int \!\! d^4x'\sqrt{-g(x')}~j(x')\nonumber \\
        &\times& \!\! \left[
W_\mu^{(-,P,\omega,k_\perp)}(x)W^\mu_{(-,P',\omega',k'_\perp)}(x)\overline{W_\nu^{(-,P,\omega,k_\perp)}(x')}~\overline{W^\nu_{(-,P',\omega',k'_\perp)}(x')}\right. \nonumber \\ 
&+& \!\! W_\mu^{(-,P,\omega,k_\perp)}(x)W^\mu_{(+,P',\omega',k'_\perp)}(x)\overline{W_\nu^{(-,P,\omega,k_\perp)}(x')}\overline{W^\nu_{(+,P',\omega',k'_\perp)}(x')} \nonumber \\
&+& \!\! W_\mu^{(+,P,\omega,k_\perp)}(x)W^\mu_{(-,P',\omega',k'_\perp)}(x)\overline{W_\nu^{(+,P,\omega,k_\perp)}(x')}\overline{W^\nu_{(-,P',\omega',k'_\perp)}(x')} \nonumber \\
&+& \!\! \left. W_\mu^{(+,P,\omega,k_\perp)}(x)W^\mu_{(+,P',\omega',k'_\perp)}(x)\overline{W_\nu^{(+,P,\omega,k_\perp)}(x')}\overline{W^\nu_{(+,P',\omega',k'_\perp)}(x')} \right], \nonumber 
\end{eqnarray}
where $\mathcal{I}\equiv\{P,\omega,k_\perp\}$ and $d\mu\equiv d\omega \,d^2\textbf{k}_\perp$.

\begin{widetext}
Next, let us assume that the current $j(x)$ has support in the right Rindler wedge. In this case, using the Unruh modes~\eqref{Unruh modes} and \eqref{Unruh modes 2}, the emission probability reads

\begin{eqnarray}
\label{two-photon probability}
        P_M^{E}
        &=&\sumint_{_\mathcal{I}} d\mu\sumint_{_{\mathcal{I}'}} d\mu'
        \left[1+n(\omega)\right]\left[1+n(\omega')\right]\left|2\int d^4x\sqrt{-g}~j~\overline{A_\nu^{R(P',\omega',k'_\perp)}}~\overline{A^\nu_{R(P,\omega,k_\perp)}}\right|^2
        \nonumber \\
        &+&\sumint_{_\mathcal{I}} d\mu\sumint_{_{\mathcal{I}'}} d\mu'~n(\omega)n(\omega')\left|2\int d^4x\sqrt{-g}~j~ A^{R(P',\omega',k'_\perp)}_\mu~ A_{R(P,\omega,k_\perp)}^\mu \right|^2
        \nonumber\\
        &+&\sumint_{_\mathcal{I}} d\mu\sumint_{_{\mathcal{I}'}} d\mu'~n(\omega)\left[1+n(\omega')\right]\left|2\int d^4x\sqrt{-g}~j~A^{R(P,\omega,k_\perp)}_\mu ~ \overline{A_{R(P',\omega',k'_\perp)}^\mu}\right|^2
        \nonumber \\
        &+&\sumint_{_\mathcal{I}} d\mu\sumint_{_{\mathcal{I}'}} d\mu'~n(\omega')\left[1+n(\omega)\right]\left|2\int d^4x\sqrt{-g}~j~ A^{R(P',\omega',k'_\perp)}_\mu~ \overline{A_{R(P,\omega,k_\perp)}^\mu}\right|^2.
\end{eqnarray}
\end{widetext}
\vspace{-1em}
To interpret the emission of Minkowski pairs according to Rindler observers, we define the amplitudes corresponding to the absorption and emission of two Rindler photons, as well as the scattering of a Rindler photon by the charge:
\begin{eqnarray*}
    \label{absorption amplitude}
    ^R\mathscr{A}^{E,\text{abs}}_{P,\omega,k_\perp;P'\omega',k'_\perp}&=&i\bra{0_R}\hat{S}_I\ket{P,\omega,k_\perp;P',\omega',k'_\perp}\nonumber\\
        &=&-2i\int d^4x\sqrt{-g}j~A_\mu^{R(P',\omega',k'_\perp)}A^\mu_{R(P,\omega,k_\perp)},
\end{eqnarray*}

\begin{eqnarray*}
    \label{emission amplitude} {}^R\mathscr{A}^{E,\text{em}}_{P,\omega,k_\perp;P',\omega',k'_\perp}&=&i\bra{P,\omega,k_\perp;P',\omega',k'_\perp}\hat{S}_I\ket{0_R}\nonumber \\
        &=&-2i\int d^4x\sqrt{-g}j~\overline{A_\mu^{R(P',\omega',k'_\perp)}}\overline{A^\mu_{R(P,\omega,k_\perp)}},
\end{eqnarray*}
\begin{eqnarray*}
    \label{scattering amplitude}{}^R\mathscr{A}^{E,\text{scatt}}_{P,\omega,k_\perp;P',\omega',k'_\perp}&=&i\bra{P,\omega,k_\perp}\hat{S}_I\ket{P',\omega',k'_\perp}\nonumber \\
        &=&-2i\int d^4x\sqrt{-g}j~A_\mu^{R(P',\omega',k'_\perp)}\overline{A^\mu_{R(P,\omega,k_\perp)}}.
\end{eqnarray*}
Using the above equations, Eq.~\eqref{two-photon probability} reads
\begin{eqnarray}
\label{inertial probability}
        P_M^E 
        &=&\sumint_{_\mathcal{I}} d\mu\sumint_{_{\mathcal{I}'}} d\mu'
        \left[1+n(\omega)\right]\left[1+n(\omega')\right]\left|{}^R\mathscr{A}^{E,\text{em}}_{P,\omega,k_\perp,P',\omega',k'_\perp}\right|^2
        \nonumber \\
        &+&\sumint_{_\mathcal{I}} d\mu\sumint_{_{\mathcal{I}'}} d\mu'~n(\omega)n(\omega')\left|{}^R\mathscr{A}^{E,\text{abs}}_{P,\omega,k_\perp,P',\omega',k'_\perp}\right|^2
        \nonumber \\
        &+&\sumint_{_\mathcal{I}} d\mu\sumint_{_{\mathcal{I}'}} d\mu'~n(\omega)\left[1+n(\omega')\right]\left|{}^R\mathscr{A}^{E,\text{scatt}}_{P',\omega',k'_\perp;P,\omega,k_\perp}\right|^2
        \nonumber \\
        &+&\sumint_{_\mathcal{I}} d\mu\sumint_{_{\mathcal{I}'}} d\mu'~n(\omega')\left[1+n(\omega)\right]\left|{}^R\mathscr{A}^{E,\text{scatt}}_{P,\omega,k_\perp;P',\omega',k'_\perp}\right|^2.
\end{eqnarray}
Thus, the two-photon emission with transverse momenta $k_\perp$ and  $k'_\perp$ in the usual vacuum of inertial observers [left-hand side of  Eq.~\eqref{scalar inertial probability}] corresponds, in general, according to Rindler observers, either to the absorption (emission) of two Rindler photons from (to) the Unruh thermal bath with the same transverse momenta $k_\perp$ and  $k'_\perp$, or to the scattering of a Rindler photon from the Unruh thermal bath with $k_\perp$ to $k'_\perp$ (or $k'_\perp$ to $k_\perp$). By disregarding back-reaction effects on the charge, the scattering process in the Rindler frame can be named after Thomson. 

Let us now connect our general result~\eqref{inertial probability} with Refs.~\cite{S06,S08}. We begin by recovering their amplitude for the emission of pairs of Minkowski photons. To do so, we use the Minkowski decomposition~\eqref{Minkowski decomposition} to express Eq.~\eqref{two state initial} as
\begin{equation}
    \ket{2}=\sumint_{_\mathcal{I}} d\mu\sumint_{_\mathcal{I'}} d\mu'\mathscr{A}_{\lambda,\textbf{k};\lambda',\textbf{k}'} \ket{\lambda,\textbf{k}; \lambda',\textbf{k}'},
\end{equation}
where
\begin{equation}
\label{Inertial EM amplitude}    \mathscr{A}_{\lambda,\textbf{k};\lambda',\textbf{k}'} \equiv -i\int d^4x \sqrt{-g}~j(x)\frac{\epsilon_\mu(\lambda,\textbf{k})\epsilon^\mu(\lambda',\textbf{k}')}{16\pi^3\sqrt{kk'}}e^{ik_\mu x^\mu}e^{ik'_\nu x^\nu}.
\end{equation}
Here,~\cite{Matsas96}
\begin{equation}
    j(x') = \frac{e^2}{2m}\frac{\delta^{(3)}\left[\textbf{x}'-\textbf{x}(\tau)\right]}{\sqrt{-g(x')}\,u^0},
\end{equation}
where $\tau$ is the charge's proper time and $u^0=dx^0/d\tau$.
In Ref~\cite{S06}'s notation, $u^0=1/\sqrt{1-\dot{\textbf{r}}_e^2\,}$ with  $``\;\dot{}\;" \equiv d/dt$, $\textbf{r}_e\equiv \textbf{x}(\tau)$, leading to 
    \begin{eqnarray}
\label{Eq: inertial frame amplitude}
        \mathscr{A}_{\lambda,\textbf{k};\lambda',\textbf{k}'} &=&-i\frac{e^2}{2m}\frac{\epsilon_\mu(\lambda,\textbf{k})\epsilon^\mu(\lambda',\textbf{k}')}{16\pi^3\sqrt{kk'}}\int dt~ \sqrt{1-\dot{\textbf{r}}_e^2
        }
        \nonumber \\
        &\times& e^{i(k+k')t-i(\textbf{k}+\textbf{k}')\cdot\textbf{r}_e(t)},
    \end{eqnarray}
which agrees with the two-photon amplitude of Ref.~\cite{S06} corrected by a factor of $2$ (in line with Ref.~\cite{S08}). Assuming Thomson scattering by a non-relativistic electron in the laboratory frame, $\dot{\textbf{r}}^2_e\ll 1$, Eq.~\eqref{Eq: inertial frame amplitude} renders the amplitude of Ref.~\cite{S08}. (Equation~\eqref{Eq: inertial frame amplitude} and the corresponding ones in Refs.~\cite{S06,S08} only differ from each other concerning the fact that momenta in the former are continuous in contrast to the latter.) As pointed out in Ref.~\cite{S06} there is a correlation between the polarizations of the emitted Minkowski photons and their momenta encoded in $\epsilon_\mu(\lambda,\textbf{k})\epsilon^\mu(\lambda',\textbf{k}')$, namely, photons with $\textbf{k} \propto \textbf{k}'$ must have the same polarization $\lambda=\lambda'$.

Finally, we show that the conjecture that the two-photon emission in the inertial frame corresponds to the scattering of Rindler photons in the Unruh thermal bath (see Ref.~\cite{S06}) is valid for uniformly accelerated charges (but not in general as made clear by Eq.~\eqref{inertial probability}; see also discussion below Eq.~\eqref{Eq: abs UA}). Let us consider that the charge is uniformly accelerated along the $z$~direction with $\xi,x,y=0$. In this case, 
$$
j(x)=\mathfrak{g}\delta(\xi)\delta^{(2)}(\textbf{x}_\perp),
$$ 
where $\mathfrak{g}\equiv e^2/(2m)$ is the coupling constant. Then, the absorption amplitude calculated before becomes
\begin{equation}
    {}^R\mathscr{A}^{E,\text{abs}}_{P,\omega,k_\perp;P'\omega',k'_\perp} = 4\pi i \mathfrak{g}\mathcal{F}_{\text{abs}}(P,\omega,k_\perp;P',\omega',k'_\perp)\delta(\omega+\omega'),
\end{equation}
where
\begin{eqnarray}
    \label{Fabs}
    &&\mathcal{F}_{\text{abs}}=-F_{\omega,\textbf{k}_\perp}(0,\textbf{0})F_{\omega', \textbf{k}'_\perp}(0,\textbf{0})(k_xk'_x+k_yk'_y)\frac{\delta_{P1}\delta_{P'1}}{k_\perp k'_\perp}\nonumber \\
    &+& \left[\left(\frac{d}{d\xi}F_{\omega, \textbf{k}_\perp}(\xi,\textbf{0})\frac{d}{d\xi}F_{\omega', \textbf{k}'_\perp}(\xi,\textbf{0})\right)_{\xi\to0}\! + \omega\omega'F_{\omega, \textbf{k}_\perp}(0,\textbf{0}) \right. \nonumber \\
&\times&  F_{\omega',\textbf{k}'_\perp}(0,\textbf{0}) \Bigg] \frac{\delta_{P2}\delta_{P'2}}{k_\perp k'_\perp}.
\end{eqnarray}

Doing the same for the other amplitudes, we get
\begin{equation}
    {}^R\mathscr{A}^{E,\text{em}}_{P,\omega,k_\perp;P',\omega',k'_\perp} = 4\pi i \mathfrak{g}\mathcal{F}_{\text{em}}(P,\omega,k_\perp;P',\omega',k'_\perp)\delta(\omega+\omega')
\end{equation}
and
\begin{equation}
\label{scattering amplitude rindler ua}
    {}^R\mathscr{A}^{E,\text{scatt}}_{P,\omega,k_\perp;P',\omega',k'_\perp}=4\pi i \mathfrak{g}\mathcal{F}_{\text{scatt}}(P,\omega,k_\perp;P',\omega',k'_\perp)\delta(\omega-\omega'),
\end{equation}
where $\mathcal{F}_{\text{em}}=\mathcal{F}_{\text{abs}}$ and
\begin{eqnarray}
    \label{Gscatt}
    &&\mathcal{F}_{\text{scatt}}
    = -F_{\omega,\textbf{k}_\perp}(0,\textbf{0})F_{\omega', \textbf{k}'_\perp}(0,\textbf{0})(k_xk'_x+k_yk'_y)\frac{\delta_{P1}\delta_{P'1}}{k_\perp k'_\perp} \nonumber \\
    &+& \left[ \left(\frac{d}{d\xi}F_{\omega, \textbf{k}_\perp}(\xi,\textbf{0})\frac{d}{d\xi}F_{\omega', \textbf{k}'_\perp}(\xi,\textbf{0})\right)_{\xi\to0} -\omega\omega'F_{\omega, \textbf{k}_\perp}(0,\textbf{0}) 
    \right. \nonumber \\
    &\times& F_{\omega',\textbf{k}'_\perp}(0,\textbf{0})\bigg] \frac{\delta_{P2}\delta_{P'2}}  {k_\perp k'_\perp}.
\end{eqnarray}
We can see immediately that the absorption and emission processes will not contribute due to the presence of $\delta(\omega+\omega')$. (Note that, as in the scalar field analysis, discussed in Sec.~\ref{sec: scalar pairs case}, this result is specific to uniform acceleration.) Consequently, only the scattering process accounts for the production of Minkowski pairs in the inertial frame. From Eq.~\eqref{inertial probability}, we have
\begin{equation}
    P_M^E = 2P^{E,\text{scatt}}_R,
\end{equation}
where
\begin{equation}
    \label{Eq: Scatt probability}
    P^{E,\text{scatt}}_R = \sumint_{_\mathcal{I}} \! d\mu\sumint_{_{\mathcal{I}'}} \! d\mu' n(\omega)\left[1+n(\omega')\right] \left|{}^R\mathscr{A}^{E,\text{scatt}}_{P,\omega,k_\perp;P',\omega',k'_\perp}\right|^2
    \end{equation}
and we recall that $d\mu\equiv d^2\textbf{k}_\perp d\omega $.  It is also interesting to note from Eq.~\eqref{Gscatt} that there is no crossed scattering, leading Rindler photons with polarization $P=1$ into $P=2$ and vice versa.

\begin{figure}[H]
    \centering
    \includegraphics[width=1\linewidth]{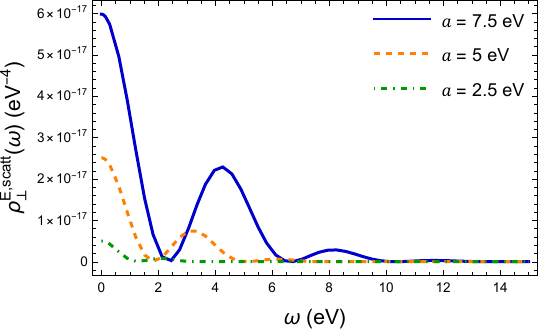}
    \caption{Energy distribution with fixed transverse momenta for the Rindler description of the electromagnetic case for different values of the electron's proper acceleration. Here, we have assumed $k_\perp=1$ eV and $k_\perp'=0.5$ eV. Note that $\rho_\perp^{E, \mathrm{scatt}}$ is symmetric under exchange of $k_\perp$ and $k'_\perp$ (see Eq.~\eqref{Gscatt}).}
    \label{fig:Energy distribution for different values of k}
\end{figure}

It is again instructive to see the energy distribution of the scattered Rindler photons. To do this, let us evaluate the scattering probability with fixed transverse momenta $\textbf{k}_\perp$ and $\textbf{k}'_\perp$ from Eq.~\eqref{Eq: Scatt probability}. By using Eq.~\eqref{scattering amplitude rindler ua}, we have
\begin{eqnarray}
        P_{\perp}^{E,\text{scatt}} \!\! & \equiv&  \!\! \frac{dP_{R}^{E, \mathrm{scatt}}}{d^2\textbf{k}_\perp d^2\textbf{k}_\perp'} \nonumber \\
        \!\!&=&\!\! \frac{4e^4\pi^2}{m^2}\int_0^\infty \!\!\!\! d\omega\int_0^{\infty} \!\!\!\! d\omega' \!\! \sum_{P,P'=1}^{2}\left|\mathcal{F}_{\text{scatt}}(P,\omega,k_\perp; P',\omega',k'_\perp)\right|^2\nonumber \\
        \!\! &\times& \!\! n(\omega')\left[1+n(\omega)\right]\delta^2(\omega-\omega'),
\end{eqnarray}
where we recall that $\mathfrak{g}=e^2/(2m)$. The integral over $\omega'$ can be easily performed, yielding
\begin{equation}
\label{transversal probability over time}
   \Gamma^{E,\text{scatt} }_\perp\equiv\frac{P_{\perp}^{E,\text{scatt}}}{T_{\rm{tot}}}=\int_0^\infty d\omega~\rho^{E,\text{scatt}}_\perp(\omega),
\end{equation}
where, we recall that $T_{\rm{tot}}= 2\pi\delta(0)$ is the total proper time, and
\begin{equation}
    \rho^{E,\text{scatt}}_\perp\equiv \frac{e^4\pi}{2m^2}\sum_{P=1}^{2}\sum_{P'=1}^{2} \frac{\left|\mathcal{F}_{\text{scatt}}(P, \omega,k_\perp;P',\omega,k'_\perp)\right|^2}{\sinh^2{\left(\pi\omega/a\right)}}.
\end{equation}
The plot of $\rho_\perp^{E,\mathrm{scatt}}$ for $k_\perp=1$ eV and $k'_\perp=0.5$ eV is shown in Fig.~\ref{fig:Energy distribution for different values of k}. (The values of $a$ and $k_\perp$ were chosen according to the characteristic scales achievable with non-relativistic optical lasers operating at a frequency of a few eV.)  Note that in both cases the scattering of soft Rindler particles is favored over high-frequency ones, as in the scalar case (see Fig.~\ref{fig:Rindler plot}).

\begin{figure}[H]
    \centering
    \includegraphics[width=1\linewidth]{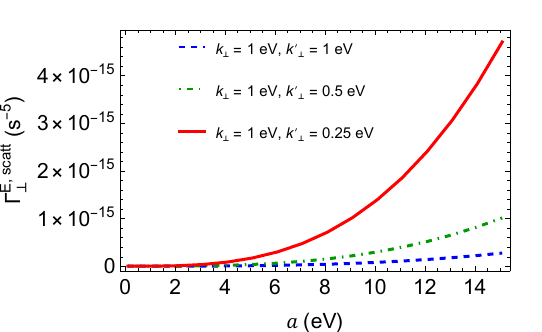}
    \caption{Scattering rate of Rindler photons (with fixed transverse momenta) from the Unruh thermal bath by a uniformly accelerated electron as a function of the proper acceleration.}
    \label{fig:Transversal probability Rindler EM}
\end{figure}

Lastly, we can numerically solve Eq.~\eqref{transversal probability over time} and plot the scattering rate $\Gamma^{E,\text{scatt}}_\perp$ as a function of the proper acceleration $a$ (see Fig.~\ref{fig:Transversal probability Rindler EM}). Note that, for the same proper acceleration, photons emitted with small values of $k_\perp$ is favored over higher ones. It can also be shown that $\Gamma_\perp^{E,\mathrm{scatt}}$ scales with $a^3$, as in the scalar case (see Eq.~\eqref{Eq: scalar inertial rate}).

We emphasize that we have considered the scattering rate of Rindler photons per fixed transverse momenta~$\Gamma_\perp^{E,\mathrm{scatt}}$ rather than the total scattering rate as the latter exhibits an infrared divergence in $k_\perp$, in contrast to the scalar case (see Eq.~\eqref{Equação 49}). This is similar to what occurs in the Larmor radiation case~\cite{M08}.

   \section{Discussion and closing remarks }
   \label{sec: Final}

The Unruh effect plays a crucial role in ensuring the internal consistency of quantum field theory in uniformly accelerated frames~\cite{UW84}. Yet, its physical reality remains contested, largely because part of the scientific community considers that no satisfactory experimental evidence has been achieved up to the moment. In this work, we have examined the emission of pairs of low-energy Minkowski photons by an accelerated electron and how it is perceived in the Rindler frame. (By low-energy we mean photons with energies much smaller than the electron's mass.) We show that the emission of pairs of Minkowski photons corresponds, in general, to the incoherent combination of three distinct processes  according to Rindler observers: scattering, and emission and absorption of pairs of Rindler photons. In the special case of uniformly accelerated charges, the radiation observed in the inertial frame can be fully accounted for the scattering channel in the Rindler frame, as suggested in Refs.~\cite{S06,S08}. We have also noted that arbitrarily soft Rindler photons still give a relevant contribution in the uniformly accelerated case. These findings highlight that the observation in the laboratory of the emission of pairs of Minkowski photons by accelerated charges can be seen as an experimental evidence supporting the Unruh effect.

\acknowledgments

   J.V.O.P. was fully supported by the São Paulo Research Foundation (FAPESP) under grant 2024/18601-5. R.B. and G.G. were partially supported by EPSRC Grants No. EP/X010791/1 and No. EP/X01133X/1. G.G. is also a member of the Quantum Sensing for the Hidden Sector (QSHS) Collaboration, supported by STFC Grant No. ST/T006277/1. G.E.A.M. was partially supported by the National Council for Scientific and Technological Development and FAPESP under Grants No. 301508/2022-4 and No. 2022/10561-9, respectively.

\section*{DATA AVAILABILITY}
The data that support the findings of this article are openly available~\cite{Data}.

\appendix
\section{Interaction action~\eqref{interaction action} from QED physics}
\label{Appendix}
\begin{figure}[h!]
    \centering
    \includegraphics[width=0.75\linewidth]{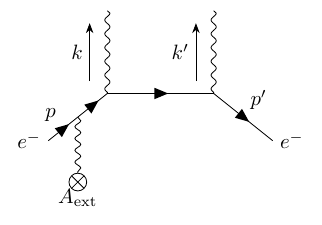}
    \caption{Feynman diagram for an electron emitting two photons. (The external line representing the accelerating field is not taken into account in the amplitude calculation, since in our semiclassical approach this is encoded in the choice of $j(x)$.)}
    \label{fig:FeynmanDiagram}
\end{figure}
In order to justify the interaction action~\eqref{interaction action}, let us consider the emission of two photons by an electron under the framework of quantum electrodynamics (QED). Let $p$ and $p'$ be the initial and final four-momenta of the electron, respectively, and $k$ and $k'$ the four-momenta of the emitted photons, see Fig.~\ref{fig:FeynmanDiagram}. Thus, the amplitude for such a process is given by~\cite{Peskin}
\begin{equation}
    i\mathcal{M} = \bar{u}(p')(-ie\gamma^\mu)\epsilon^*_\mu(k')S_F(p-k)(-ie\gamma^\nu)\epsilon^*_\nu(k)u(p),
\end{equation}
where $\epsilon_\mu$ stands for the photon polarization vectors, $u(p)$ and $\bar{u}(p')$ are spinor solutions of the Dirac equation, and $$S_F(p-k)=\frac{i(\slashed{p}-\slashed{k}+m)}{(p-k)^2-m^2}.$$ Using now that $k^2=0$ and $p^2=m^2$ (since $p$ is on-shell), we get 
\begin{equation}
    i\mathcal{M}=-e^2\bar{u}(p')\slashed{\epsilon}(k')\,\frac{i(\slashed{p}-\slashed{k}+m)}{2p_\mu k^\mu}\,\slashed{\epsilon}(k)u(p),
\end{equation}
where $\slashed{\epsilon}(k)=\gamma^\mu\epsilon_\mu^*(k)$ and similarly for $\slashed{\epsilon}(k').$ 

Let us proceed with the calculations in the frame where $p^\mu = (m,\mathbf{0})$, which yields $2p_\mu k^\mu =2mk^0$. In addition, let us assume that the energy of the emitted photons is much smaller than the electron's mass, leading us to neglect the electron recoil, \textit{i.e.}, $p'\approx p$. Hence,
\begin{equation}
\label{A3}
    i\mathcal{M} = -i\frac{e^2}{2mk^0}\bar{u}(p)\slashed{\epsilon}(k')(\slashed{p}-\slashed{k}+m)\slashed{\epsilon}(k)u(p).
\end{equation}
Next, from the Dirac equation, $ \bar{u}(p)(\slashed{p}-m)=0$, we find
\begin{equation}
\label{A4}
    i\mathcal{M} =-i\frac{e^2}{2mk^0}\bar{u}(p)\left(\slashed{\epsilon}(k')\slashed{p}-\slashed{\epsilon}(k')\slashed{k}+\slashed{p}\slashed{\epsilon}(k')\right)\slashed{\epsilon}(k)u(p).
\end{equation}
Now, from the Clifford algebra $\slashed{\epsilon}(k')\slashed{p}+\slashed{p}\slashed{\epsilon}(k')=2p_\mu\epsilon^\mu(k')$. Assuming the \textit{radiation gauge}, $\epsilon_\mu(k')$ has no timelike component; so $p_\mu\epsilon^\mu(k')=0.$ Combining these results, Eq.~\eqref{A4} reads
\begin{eqnarray}
    i\mathcal{M} &=&i\frac{e^2}{2mk^0}\bar{u}(p)\slashed{\epsilon}(k')\slashed{k}\slashed{\epsilon}(k)u(p)\nonumber \\
    &=&i\frac{e^2}{2mk^0}\epsilon_\mu(k')k_\nu\epsilon_\rho(k)\bar{u}(p)\gamma^\mu\gamma^\nu\gamma^\rho u(p).
\end{eqnarray}
Using now $$\gamma^\mu\gamma^\nu\gamma^\rho=\eta^{\mu\nu}\gamma^\rho-\eta^{\mu\rho}\gamma^\nu+\eta^\nu\rho\gamma^\mu+i\epsilon^{\mu\nu\rho\sigma}\gamma_5\gamma_\sigma,$$ and neglecting contributions from the electron's spin, we drop the last (spin-coupling) term, getting
\begin{eqnarray}
\label{A6}
    i\mathcal{M} &=& i\frac{e^2}{2mk^0}\left[\epsilon_\mu(k')k^\mu\epsilon_\rho(k)\bar{u}(p)\gamma^\rho u(p)\right.\nonumber\\
    &-&\epsilon_\mu(k')\epsilon^\mu(k)k_\nu\bar{u}(p)\gamma^\nu u(p)\nonumber\\ &+&\left.\epsilon_\mu(k')k_\nu\epsilon^\nu(k)\bar{u}(p)\gamma^\mu u(p)\right].
\end{eqnarray}
The third term vanishes because $k_\nu\epsilon^\nu(k)=0$. For the first and second terms, we use the Gordon relation
$$\bar{u}(p')\gamma^\mu u(p)=\bar{u}(p')\left(\frac{p'^\mu+p^\mu+i\sigma^{\mu\nu}(p'-p)_\nu}{2m}\right) u(p),$$
which reads $$\bar{u}(p)\gamma^\mu u(p)=\frac{p^\mu}{m} \bar{u}(p)u(p)=\frac{p^\mu}{m}$$ for $p'=p$ and $\bar{u}(p)u(p)=1$.
Applying this in Eq.~\eqref{A6}, we obtain
\begin{equation}
    i\mathcal{M} =i\frac{e^2}{2mk^0}\left(\epsilon_\mu(k')k^\mu\epsilon_\nu(k)\frac{p^\nu}{m}-\epsilon_\mu(k')\epsilon^\mu(k)k_\nu\frac{p^\nu}{m}\right).
\end{equation}
Again, the first term vanishes due to $\epsilon_\nu(k)p^\nu=0$, and we end up with
\begin{equation}
    i\mathcal{M} =-i\frac{e^2}{2m}\epsilon_\mu(k')\epsilon^\mu(k),
\end{equation}
which has the same structure as the amplitude computed in Eq.~\eqref{Inertial EM amplitude}. Therefore, we can extract the same QED physics at low energies from the effective action~\eqref{interaction action}.

\end{document}